\documentclass[apj]{emulateapj}
\usepackage{apjfonts}

\begin{document}
\title{Migration and growth of giant planets in self-gravitating disks with varied thermodynamics}
\author{Laure Fouchet$^{1}$ \& Lucio Mayer$^{2,1}$}
\footnotetext[1]{Department of Physics, Institut f\"ur Astronomie, ETH Z\"urich, Schaffmattrasse 10, CH-8093 Z\"urich, Switzerland,  fouchet@phys.ethz.ch}
\footnotetext[2]{Institut f\"ur Theoretische Physik, University of Z\"urich, Winterthurerstrasse 190, CH-8057 Z\"urich, Switzerland}

\begin{abstract}

We report on the results of novel global high-resolution three-dimensional simulations of disk-planet interaction  
which incorporate simultaneously realistic radiation physics and the self-gravity of the gas, as well as allowing the planet to move.
We find that thermodynamics and radiative physics have a remarkable effect on both migration and accretion of Jupiter mass planets.
In simulations with radiative transfer adopting flux-limited diffusion, inward migration can be
decreased by about $30\%$ relative to the isothermal case, while in adiabatic runs migration
nearly shuts off after a few tens of orbits. Migration varies
because the relative strength of the inner and outer spiral perturbations is affected by thermodynamics, thus changing
the net torque acting on the planet.
Mass accretion rates on the planet can be reduced by 
more than an order of magnitude going from isothermal to radiative transfer and adiabatic simulations.
A circumplanetary disk always forms except in adiabatic runs. With radiative transfer the disk is sub-keplerian
($v_{rot}/v_{kep} \sim 0.7$) owing to significant pressure support. We
discuss the effect of circumplanetary disk structure on the drift of embedded dust grains
and planetesimals and thus on the formation of the rocky satellites of giant planets.
\end{abstract}


\section{Introduction}

In the conventional picture, giant planets form via a two-stage process, known as core-accretion. In such a model, first a massive rocky
core is assembled via gravitational accumulation of planetesimals in the protoplanetary disk and then the core begins to accrete 
the surrounding gas once it has grown massive enough \citep{Pollack96,Ida04,Alibert05}. 
The growing planet excites density waves as it moves through the disk. Such waves exert a torque
on the planet \citep{Lin93} whose net effect 
is to extract angular momentum from its orbit  for standard protosolar nebula models  \citep{Ward97}.


Different regimes of migration have been identified by numerous studies depending on the planet's mass and on the
importance of co-rotation torques \citep[see][for a review]{PPV}. In particular, when the planet has a mass comparable to Saturn or
larger the interaction with the disk becomes markedly nonlinear 
(type II migration) 
and the planet 
carves a gap in the disk as a result of the planetary tide \citep{Lin93,Crida07}.
Then, the planet decouples dynamically from the disk 
and migrates inward at a pace determined by the local viscous timescale
\citep{Lin86}.

A major assumption of almost all simulations published so far is that the disk is locally isothermal. This means that any heating 
is immediately radiated away, which would only be true if the disk was optically thin.
The planet moves supersonically in the disk, shock-heating the gas, and gas accretion onto the planet generates compressional heating.
With typical densities $10^{-11}$g/cm$^3$, the midplane of a minimum solar nebula disk is indeed optically thick ($\tau \approx 10$)
\citep{d'Angelo03}.
Recently, Paardekooper \& Mellema (2006, 2008, hereafter \citet{Paardekooper06,Paardekooper08}) have used both 2D and 3D adaptive mesh refinement simulations with radiative
transfer modeled via flux-limited diffusion finding that for low mass planets both inward migration and gas accretion can be strongly 
suppressed.
Klahr \& Kley (2006, hereafter \citet{KK06}) studied
Jupiter-sized planets with a static 3D grid code and flux-limited diffusion. They did not report
significant differences on migration compared with the isothermal case but found the structural
evolution of the circumplanetary gas distribution to be strongly affected by radiation physics as
noticed earlier in the 2D nested-grid calculations by \citet{d'Angelo03}.


Yet, even these recent simulations lack several important ingredients. First, they do not treat self-consistently the dynamics of the disk, 
planet and star;  the planet and the star cannot move, and in some cases a gap is introduced already at the beginning of the simulation (KK06).
Second, except for the 2D simulations of d'Angelo et al. (2003), they adopt inviscid
disks. Recently, Edgar (2007) found that giant planet migration in an isothermal viscous disk does not obey the standard type II
regime. A deep gap is never produced and hence migration does not proceed on the viscous timescale.
Finally, all these simulations neglect the self-gravity of the gas, which is required when simulating freely 
moving planets and might affect disk torques (Baruteau \& Masset 2008). Self-gravity has been previously included only in 
a few isothermal calculations \citep{Nelson03a,Nelson03b,Lufkin04}.

In this Letter we present the first high-resolution three-dimensional hydrodynamical simulations of the interaction between a massive, 
Jupiter-sized planet and a surrounding viscous protoplanetary disk that include simultaneously radiative transfer, shock heating,
self-gravity of the gas and fully self-consistent dynamics.  We study migration, mass flow towards the planet
and the circumplanetary gas distribution, exploring also the implications on the formation of satellites of giant planets.
We compare the results with those obtained for locally isothermal disks as well as other simulations with simplified 
disk thermodynamics.

\begin{table}
\scriptsize
\caption{Simulations parameters\label{tab:M}}
\begin{tabular}{rrrrrr}
\tableline
Run   & Number & Mean      & Partial   & Mean      & Partial   \\
Name  & of     & accretion & accretion & migration & migration \\
      & orbits & rate      & rate      & rate      & rate      \\
\tableline
Model M1 & M$_d$= & 0.004 M$_{Sol}$ &  &   &        \\
\tableline
%
IsoT1M1       &  39.8 & 6.34 10$^{-4}$  & 7.59 10$^{-4}$  & 1.26 10$^{-3}$ & 5.02 10$^{-4}$ \\
Adia1M1       &  80.4 & 1.02 10$^{-6}$ & 5.12 10$^{-6}$  & 7.29 10$^{-5}$ & 3.24 10$^{-4}$ \\
$\star$ IsoT200K1   & 173.4 & 5.53 10$^{-5}$  & 2.46 10$^{-4}$  & 2.87 10$^{-4}$ & 6.09 10$^{-4}$ \\
$\star$ FLD200K1    &  20.6 & 2.19 10$^{-5}$  & 2.19 10$^{-5}$  & 4.22 10$^{-4}$ & 4.22 10$^{-4}$ \\
$\star$ NIsoT200K1  &  82.3 & 0               & 0               & 6.07 10$^{-5}$ & 3.16 10$^{-4}$ \\
$\star$ Adia200K1   &  93.2 & 0               & 0               & 3.73 10$^{-5}$ & 1.83 10$^{-4}$ \\
\tableline
Model M2 & M$_d$= & 0.01 M$_{Sol}$ &    &     &      \\
\tableline
IsoT1M2       &  38.9 & 1.63 10$^{-3}$  & 1.68 10$^{-3}$  & 1.34 10$^{-3}$ & 4.01 10$^{-4}$ \\
FLD1M2        &   2.9 & 9.77 10$^{-4}$  & \nodata         & 1.15 10$^{-3}$ & \nodata        \\
NIsoT1M2      &  14.3 & 3.02 10$^{-4}$  & \nodata         & 1.22 10$^{-3}$ & \nodata        \\
Adia1M2       &  17.4 & 2.29 10$^{-5}$  & \nodata         & 9.01 10$^{-4}$ & \nodata        \\
IsoT100K2     & 167.7 & 1.27 10$^{-3}$  & 1.91 10$^{-3}$  & 6.48 10$^{-4}$ & 9.60 10$^{-4}$ \\
FLD100K2      &  27.9 & 1.14 10$^{-3}$  & 1.14 10$^{-3}$  & 8.33 10$^{-4}$ & 8.33 10$^{-4}$ \\
NIsoT100K2    & 111.8 & 0               & 6.77 10$^{-7}$ & 9.61 10$^{-5}$ & 4.00 10$^{-4}$ \\
Adia100K2     & 133.8 & 0               & 0               & 1.43 10$^{-4}$ & 2.24 10$^{-4}$ \\
\tableline
\end{tabular}
\tablenotemark{k}
\tablenotetext{k}{The Table shows the accretion and migration rates for the various
runs. The legend for the run names is as follows: IsoT refers to
Isothermal runs, FLD to runs with the Flux
  Limited Diffusion approximation, NIsoT are adiabatic runs with
  $\gamma$ = 6/5 and Adia are adiabatic runs with $\gamma$ = 7/5. 
100K corresponds to 100 000 particles, 200K to 200 000
  particles and 1M to 1 Million particles. The runs are normally done
  with a planet gravitationnal softening of R$_H$/5, where
  R$_H$=0.35~AU is the Hill radius for a Jovian planet, except for
  those marked with $\star$ for which the softening is equal to R$_H$. Accretion rates are given in M$_{J}$ yr$^{-1}$ and migration
  rates in AU yr$^{-1}$. Mean rates are computed by
  averaging over time over the full extent of the simulation. Partial rates are computed after a number of
  orbits equivalent to the maximum
  number of orbits for the FLD runs, namely after 20.6 orbits for
  model M1 and 27.9 orbits for model M2.
When the accretion rate is 0, it means that the mass
  inside the Hill radius is below the SPH mass resolution.}
\end{table}


\begin{figure}
\plotone{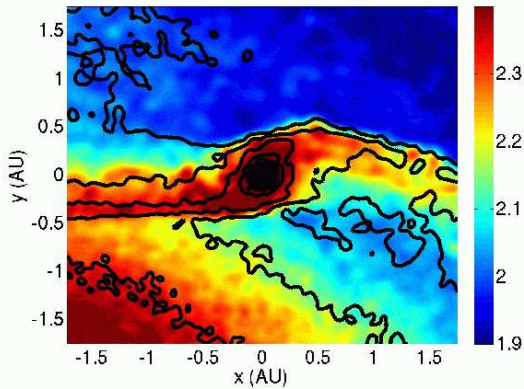}
\caption{Temperature map with overplotted midplane
  density contours of run FLD1M2 after 2.9 orbits. The shock along the spiral arms triggered by the planet is evident. The planet is seen as a hot spot in the disk.}
\label{fig:ArmFLD}
\end{figure}

\section{Initial Conditions and Simulations}

We have run different simulations with varying disk masses, resolution  and thermodynamics (see Table~1).
Our simulations include the self-gravity of the gas \citep{Nelson03a,Nelson03b,Lufkin04}.
We place a 1 Jupiter mass ($1 M_J$) planet on a circular orbit at $5$~AU for
model M1 (resp. $5.4$~AU for model M2) from a
$1 M_{\odot}$ mass star in an initially axisymmetric disk.
The planet and the star are represented by softened N-Body particles free to move in response 
to their mutual gravity and that of the disk. There is no gap initially.
 Mass is allowed to accumulate around the planet since we do not use a prescription for accretion.
Our initial disk extends from 
2 to 20~AU in model M1 and from 1 to 25~AU in model M2.
The surface density profile is $\Sigma = \Sigma(5~AU)
(r/5~AU)^{-1.5}$. We adopt an initial surface density of 
$\Sigma(5~AU)=75~\mathrm{g\ cm^{-2}}$ (resp. $150~\mathrm{g\ cm^{-2}}$),
which yield a total disk mass of $M_\mathrm{disk} = 0.004~M_{\sun}$ for model M1
and 0.01~$M_{\sun}$ for model M2.
The vertical density profile is initially gaussian.  
We assume the same disk temperature profile as in \citet{KK06} for the sake of comparison, $T=T_0 r^{-1}$ ($T$ is independent on z). 
The pressure scale-height is initially chosen to be $H/r = 0.05$ and implies $T{_0} = 102$ K. 


As for disk thermodynamics, we consider the following different cases:
(1) locally isothermal (IsoT) runs, in which
the temperature of individual particles is kept constant over time;
(2) adiabatic (Adia) and nearly isothermal (NIsoT) runs, 
in which the gas is evolved adiabatically assuming an adiabatic index of, respectively, $\gamma=7/5$ and  $\gamma=6/5$
and solving the internal energy equation including shock heating; (3)
radiative transfer (FLD) runs, in which we solve the energy
equation with added flux-limited
diffusion and blackbody radiation at the disk edge to model the
thermal energy flow in the disk, as described in \citet{Mayer07}. We adopt realistic
opacities by d'Alessio et al. (1997).

\begin{figure}
\plottwo{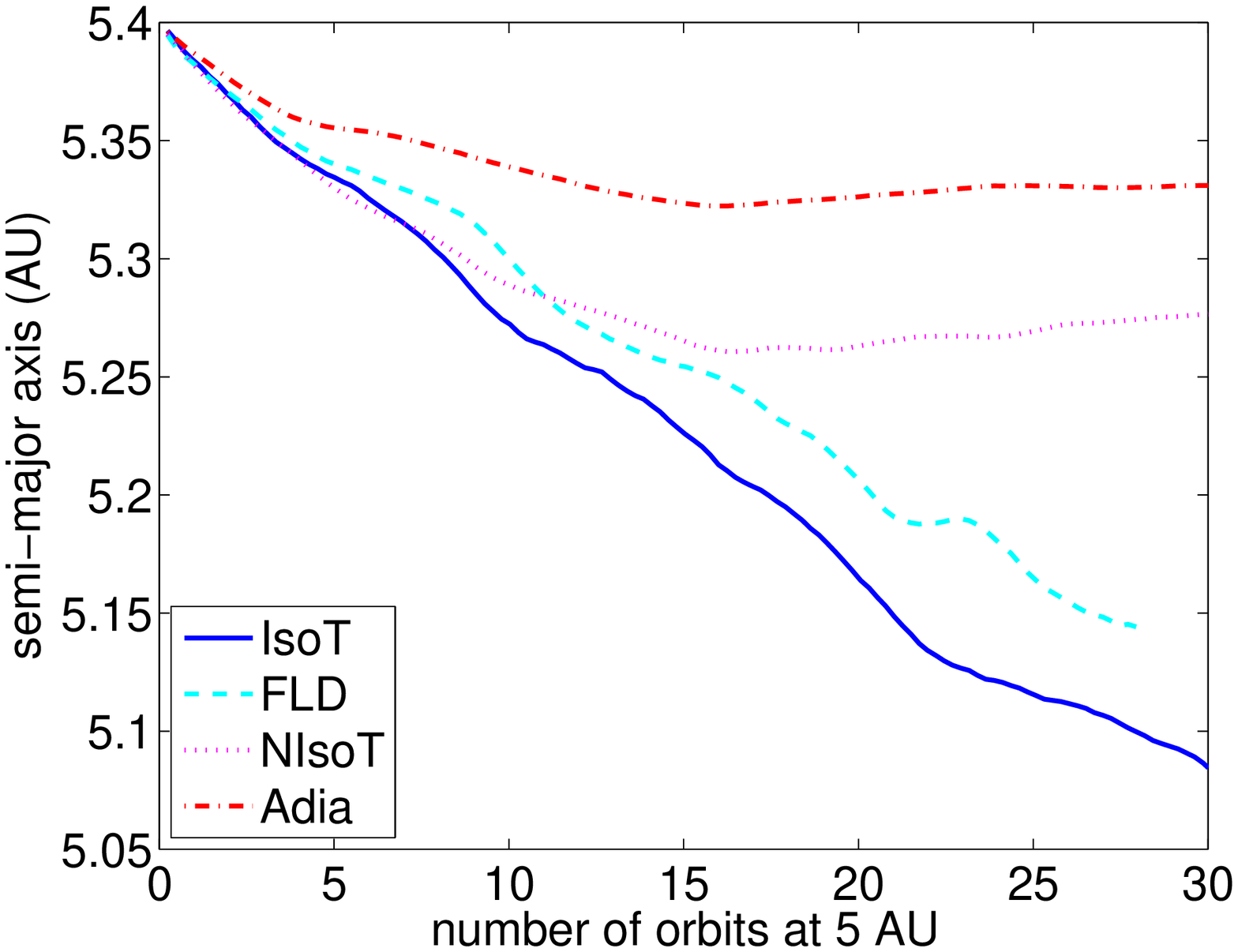}{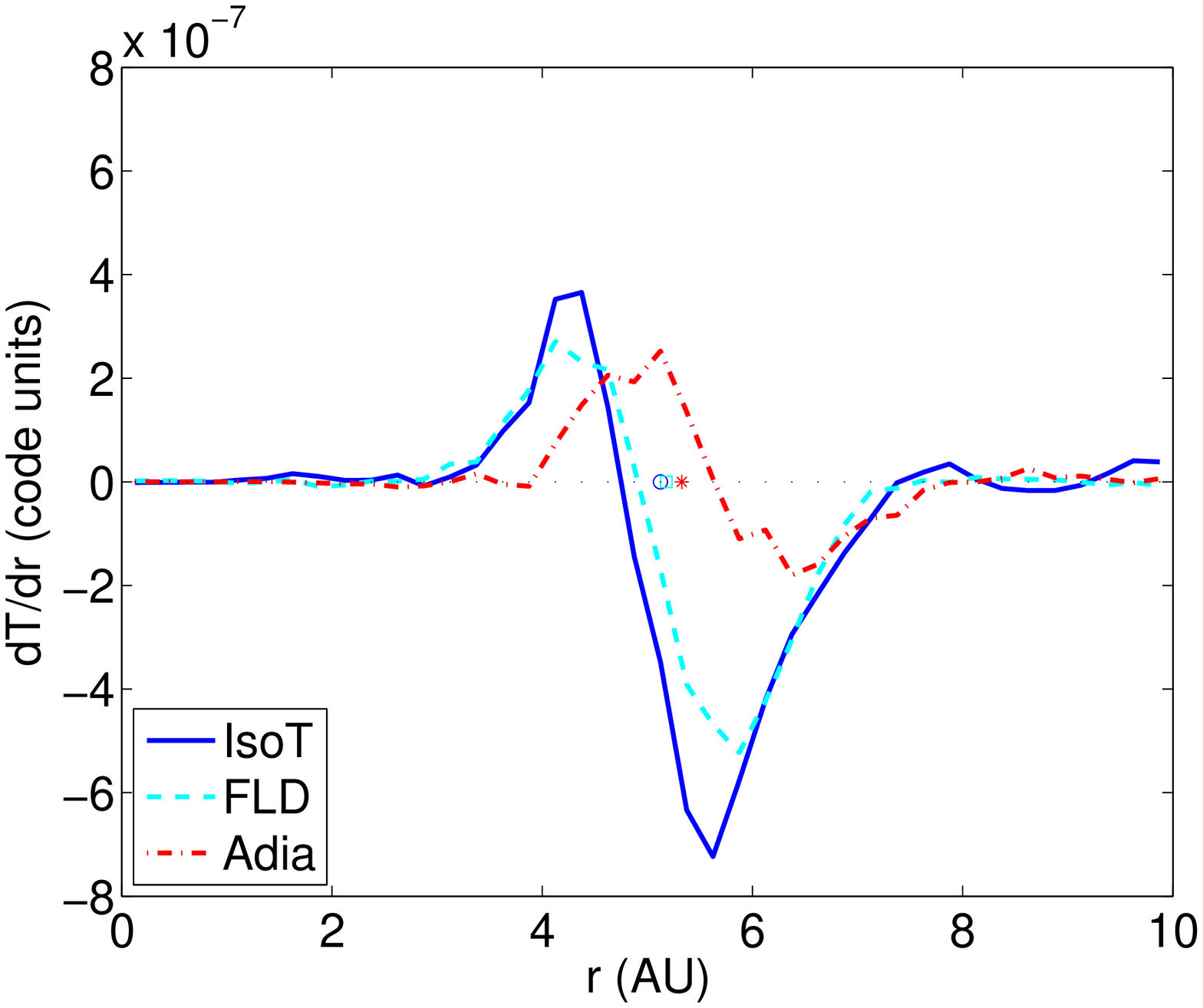}
\caption{Semi-major axis of the planet versus time (left) and
  azimuthally averaged radial torques exerted by the disk on the
  planet (right) as a function of the distance from the star for runs IsoT100K2, FLD100K2,
  NIsoT100K2 and Adia100K2 after 20 orbits. Note that, in the adiabatic case, the inner and outer torques nearly balance. The legend for the different curves is in the figure. The position of the
  planet is shown with symbols of different colors depending on the
  run (right panel).}
\label{fig:MigrTorque}
\end{figure}

The simulations are performed with the parallel 3D Tree+SPH code GASOLINE \citep{Wadsley04}, 
which includes adaptive multiple timesteps. In radiative transfer runs radiative timesteps
can become very short so that only a few tens of orbits could be
explored \citep[see also][]{Paardekooper06,Paardekooper08}.
We use the standard Monaghan viscosity with coefficients $\alpha=1$ and $\beta=2$ and with the Balsara switch 
to reduce viscosity significantly in shear flows \citep{Balsara95}. We find an average equivalent 
effective Shakura-Sunyaev $\alpha_\mathrm{SS}$ parameter
$\sim 10^{-2}$, although this varies appreciably with time and location in the disk.
For comparison, protoplanetary disks are expected to have $\alpha_\mathrm{SS}$ between $10^{-4}$ and $10^{-1}$ as
a result of MHD turbulence (Papaloizou \& Nelson 2003).\\

\section{Results}

We find that both the orbital evolution and the properties of the gas flow around the planet depend significantly on disk thermodynamics. 

\subsection{Migration}

Our reference case for the study of migration is represented by the locally isothermal runs. These have been widely used in the literature, 
although only rarely in the past self-gravity was included and/or the planet was allowed to move \citep{PPV,Bate03,Nelson03a,Nelson03b}.
The planet carves a gap quite slowly compared to simulations of inviscid disks (e.g. Nelson et al. 2003a). 
However, the drop in the surface density near the planet 
reaches more than order order of magnitude after 190 orbits,  in agreement with the results of 
other three-dimensional grid-based 
and SPH simulations of Jupiter-mass planets in disk with moderate viscosity \citep{deValBorro06}. 
At this late stage the migration timescale is $\sim 3 \times 10^4$ years, somewhat faster than the timescale
of $\sim 8 \times 10^4$ years typically inferred for Jupiter mass planets evolving in inviscid disks 
on a fixed orbit \citep[e.g.][]{PPV,KK06,Edgar07}. A speed-up of the migration rate
by a factor $2-3$ for disks with $\alpha$ viscosities $\sim 10^{-2}$
has already been reported in the literature (d'Angelo et al. 2003;
Edgar 2007). For the first few tens of orbits, when the gap is barely opened, the planet excites strong
spiral modes since it is still well coupled with the disk and the migration rate is 
much faster than in the late stage (see partial migration rate in Table~1). Self-gravity also should lead to an 
acceleration of migration before the planet has opened a gap
(Baruteau \& Masset 2008).
The mean migration rate increases with increasing disk mass (Table~1), in agreement with Edgar (2007), who also studied viscous isothermal 
disks that do not open deep gaps.

In flux-limited diffusion (FLD) runs the planet has a significant effect on the temperature structure of the disk
around it. The triggered spiral arms are associated with a shock since the planet is moving supersonically. The
temperature along the spiral arms thus rises significantly because the disk midplane is optically thick (Figure~1) an effect which 
is obviously absent in isothermal runs.
Flux-limited diffusion runs are followed up to a point when they are
still in the weak gap regime, i.e. for no more
than 30 orbits. The gap is somewhat weaker in these runs because the disk is heated by the spiral shocks in the region around the planet, becoming thicker and thus making it harder to satisfy the gap opening condition. 


Despite the shallower gap, in FLD runs migration is slower than in the isothermal case, by $30\%$ for the case of the low 
disk mass, and by nearly $15\%$ for the case of the low mass disk (Figure~2, left panel and Table~1). 
The slower migration is due to an effect already seen and 
explained in \citet{Paardekooper08} for deeply embedded, low-mass planets. It is due to the change of the balance between the outer 
and inner torque acting on the planet as the thermodynamics is varied. 
In particular, we find that, the net torque is more negative in the
isothermal runs relative to the FLD ones
(Figure~\ref{fig:MigrTorque}, right panel). The variation of torque balance is due to a variation
in the structure of the spiral perturbations excited by the planet.

Torques are even more affected by the different temperature and density evolution in the adiabatic runs (Figure~2, left panel). In these runs the gas
along the spiral shocks rises its temperature by up to a factor of 5 and adiabatically expands  away from the midplane. 
As a result, after 50 orbits the gas surface density in the disk midplane around
the planet has decreased by a factor $12.5$ in the run with $\gamma =7/5$ and
by $8.3$ in the run with $\gamma = 6/5$.  Because of the decreased gas surface density, the spiral arms around the planet weaken
considerably with time, and so do the torques exerted by them. Then, after a few tens of orbits, migration nearly shuts off (Figure~2, right panel).
The planet is never able to open a gap because the radius 
of the planet's Roche lobe is smaller than the local disk semi-thickness.


\begin{figure}[h]
\plottwo{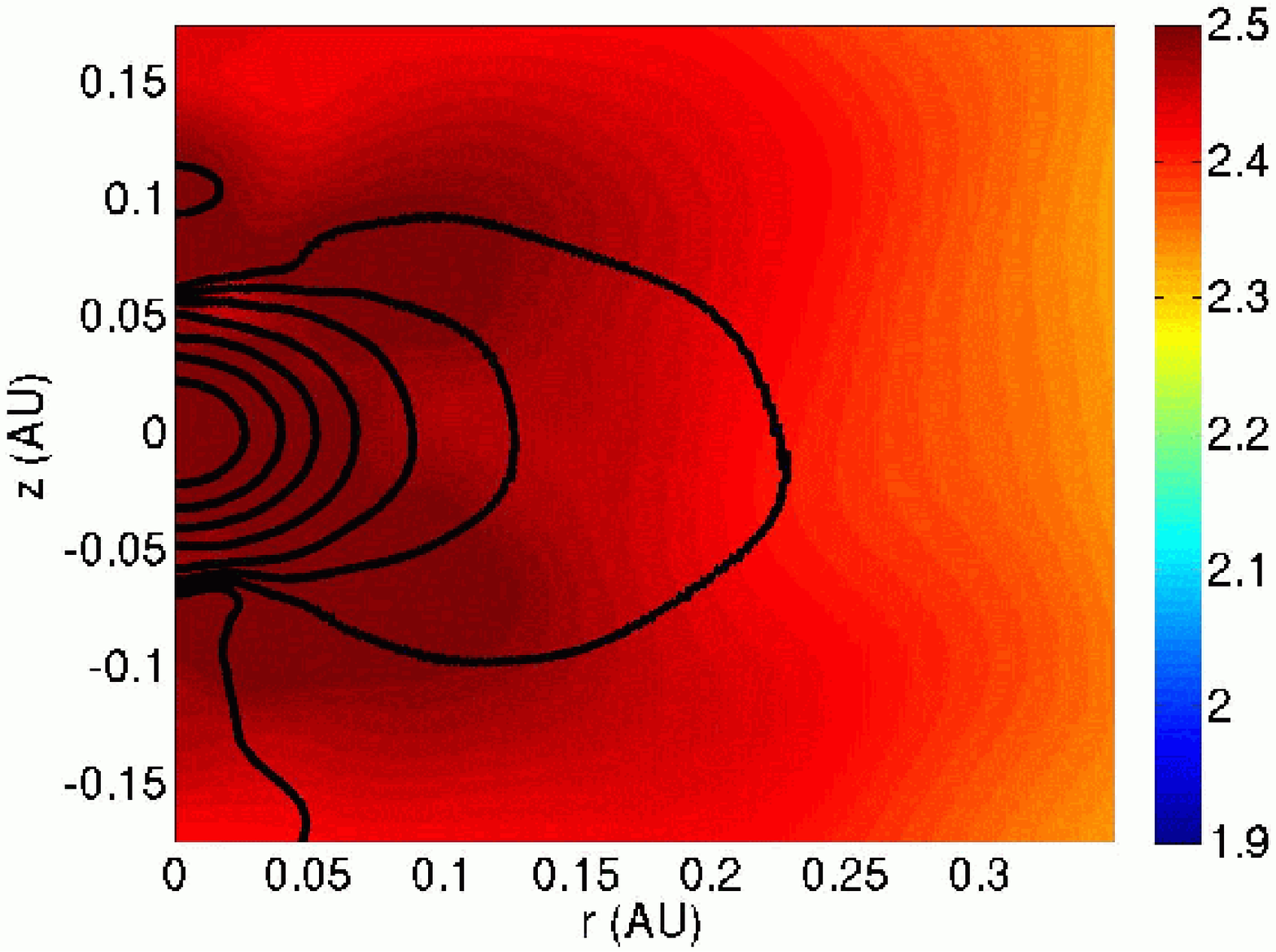}{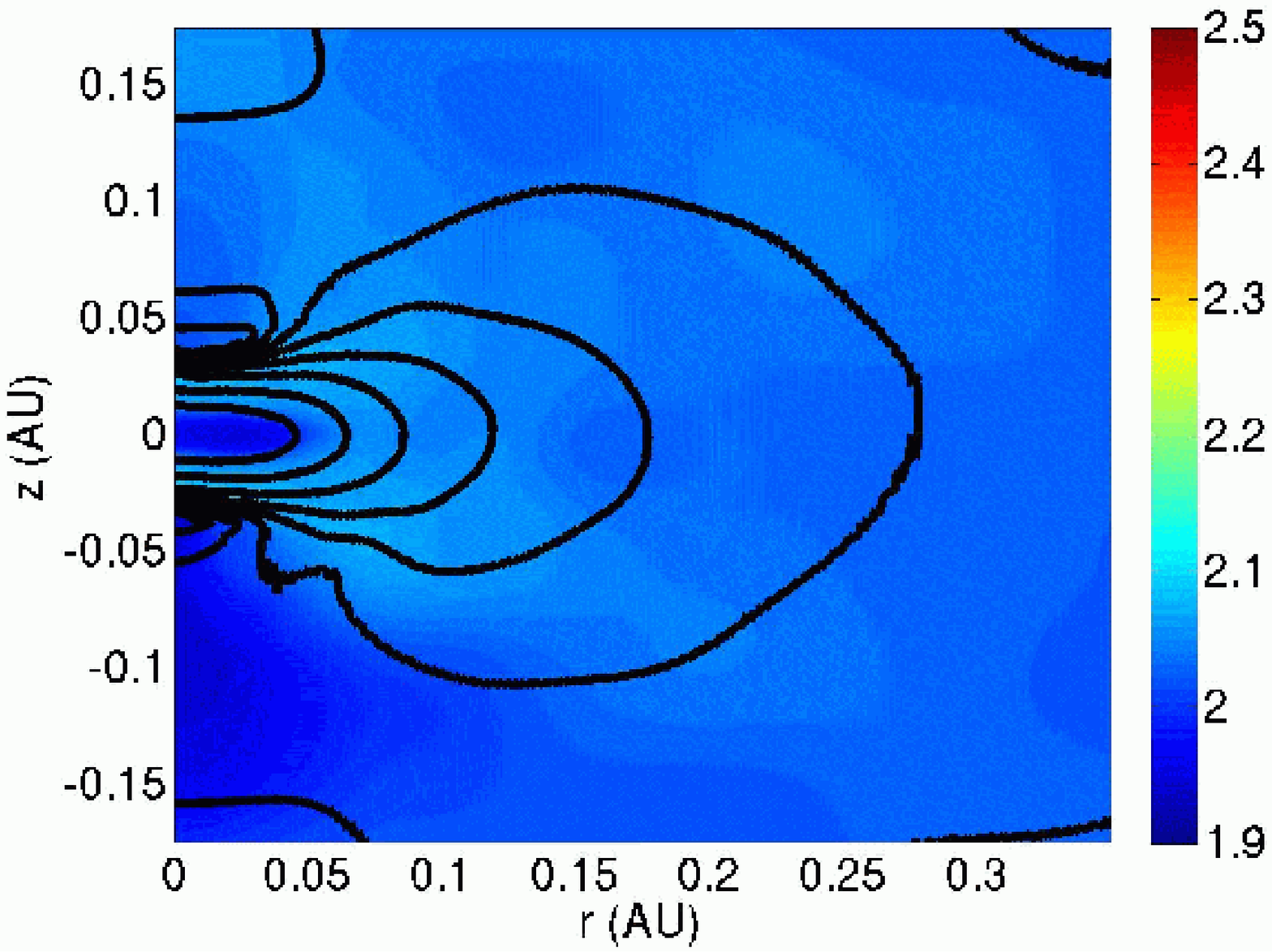}
\caption{Azimuthally averaged temperature maps of circumplanetary gas with superimposed
  density countours for FLD1M2 (left) and IsoT1M2 (right) runs after
  2.9 orbits. The colorbar gives the temperature on a logarithmic scale in K.}
\label{fig:maps}
\end{figure}

\subsection{Gas accretion and planetary growth}

Since we do not include explicitly accretion onto the planet in our simulations we calculate an hypothetical accretion rate by 
measuring the mass of gas that accumulates within the Hill radius of the planet ($R_H = 0.35$ AU) as a function of time.
The accretion rate thus defined is found to be strongly dependent on disk thermodynamics as well as on the disk mass, and,
not surprisingly, increases moderately as the gravitational softening of the planet is decreased (see Table~1). 

In isothermal M1 runs  the accretion rate averaged over nearly 200 orbits is $5 \times 10^{-5}$ M$_{J}$ yr$^{-1}$.
in agreement with previous estimates obtained for similar disk masses and effective $\alpha$ viscosities 
(e.g. Bate et al. 2003; d'Angelo et al. 2003).
\citet{KK06} compute mass accretion rates about 3 times smaller for planets
in an isothermal disk but their disk is inviscid and they start with a gap, 
both differences going in the direction of lowering somewhat the mass accretion.   

As seen in \citet{Paardekooper08} for the case of low mass planets, planetary accretion in non-isothermal disks
can be reduced compared to isothermal disks. This is because the gas is compressionally heated as it flows towards
the planet, and the resulting pressure gradient opposes further gas inflow. The temperature structure around 
the planet in one of the FLD runs is shown in Figure~\ref{fig:maps}.
For our lowest disk mass (M1) the accretion rate over the first 20 orbits is 
$\sim 2 \times 10^{-5}$ M$_{J}$ yr$^{-1}$ in the FLD run, 20 times lower
than for the isothermal during the same part of the orbital evolution (Table~1), and comparable with what 
\citet{KK06} find for their radiative transfer runs starting with or without a gap.
Similar rates were also obtained by d'Angelo et al. (2003) for their radiative viscous models.
The difference between isothermal and FLD runs is smaller in the case of the largest disk mass (M2).
In the FLD runs with in the M2 disk, which have the smallest softening, the planet should 
be able to accrete a few Jupiter masses over $\sim 10^4$ years, a time comparable to the migration rate. 







In adiabatic disks accretion, like migration, essentially shuts off after a few tens of orbits (Table~1). Due to the absence of cooling
compressional heating in this case is indeed more effective. Moreover, heating due to spiral shocks raises
the temperature of the gas even at some distance from the planet, which further stifles accretion.


\subsection{The circumplanetary disk}

A circumplanetary disk begins to form already during the first few orbits in both the isothermal and the radiative transfer runs (Figure~3).
The disk is thicker in the radiative transfer runs due to higher thermal pressure. It is sub-keplerian in both
cases. Yet, there is a fairly extended region, between  $0.15$ and $0.35$~AU from the planet, within which  
the gas azimuthal velocity is close to $70\%$ of the local keplerian velocity (Figure~4, left panel).
In the adiabatic runs the gas
becomes hot enough to form a diffuse pressure supported atmosphere around the planet rather than a dense circumplanetary
disk. In the latter case the gas is markedly sub-keplerian (Figure~4, left panel). In reality this is an extreme configuration that
could occur only if cooling is completely suppressed by some stronger heating mechanism. A protoplanetary disk strongly irradiated
by neighboring OB stars in a dense cluster might present such extreme conditions at large radii, $R > 10$ AU.
\citep{Alexander06}. 

\begin{figure}
\plottwo{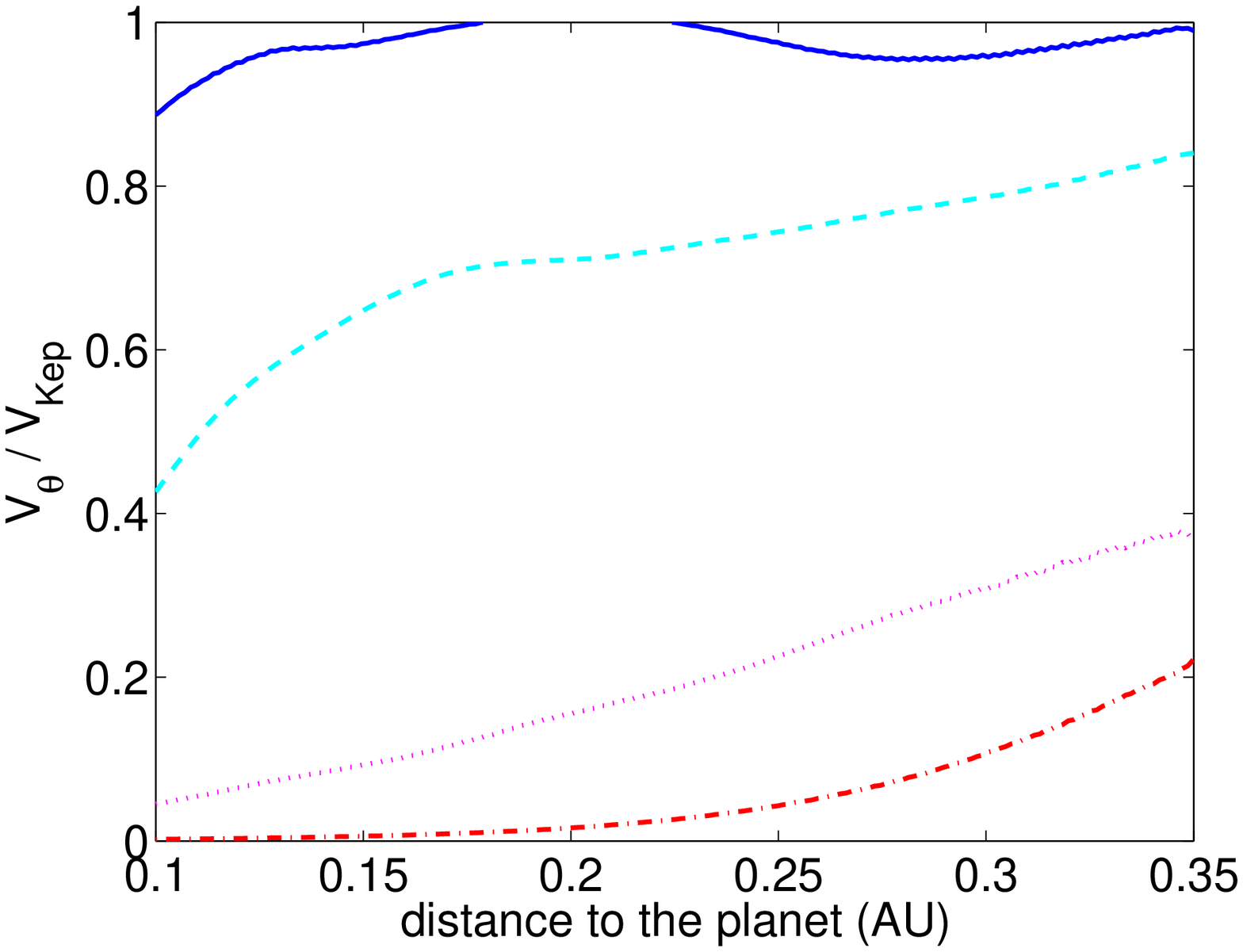}{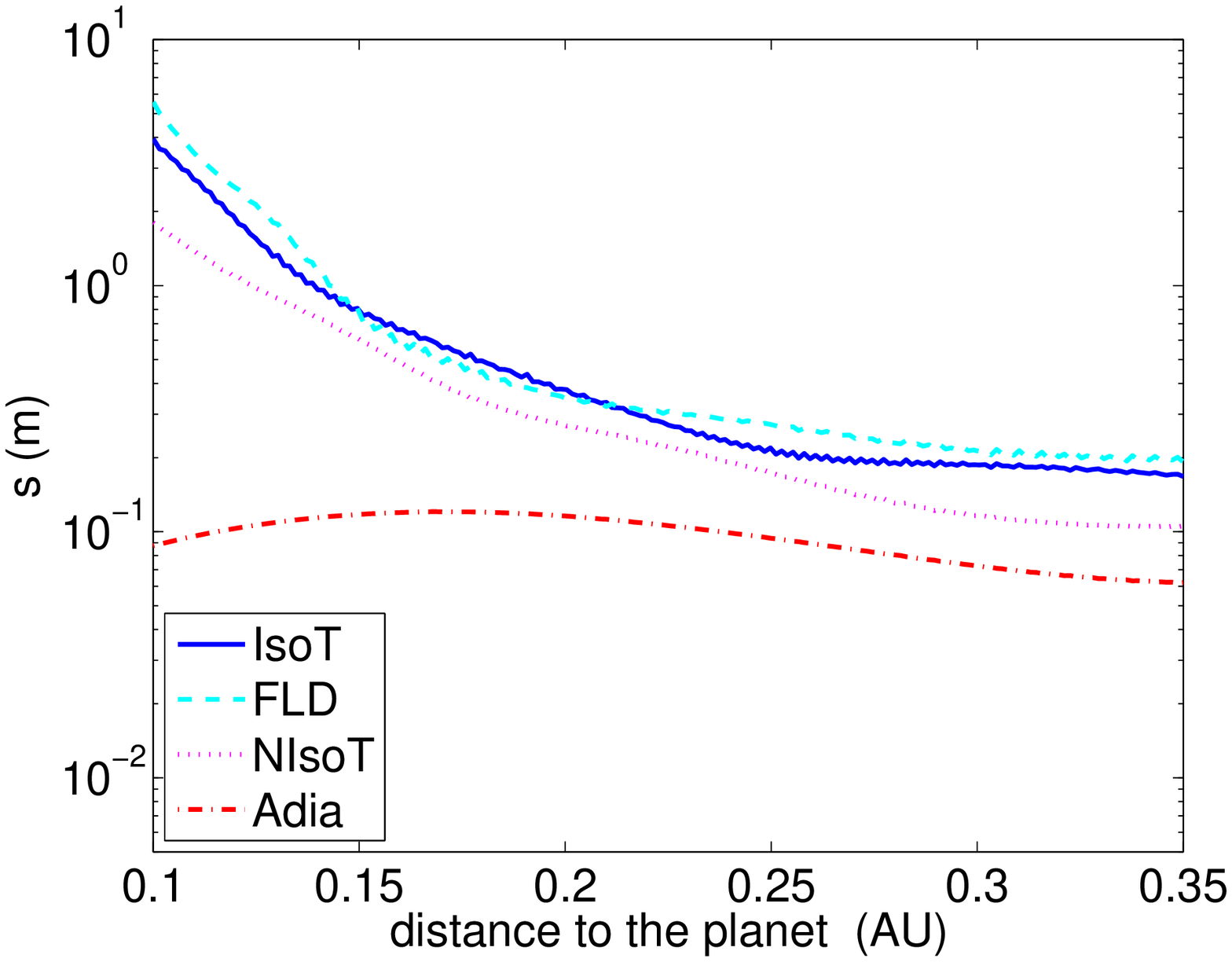}
\caption{Gas azimuthal velocity divided by the keplerian velocity
  (left) and grain size for which $T_s/T_{orb}$=1/(2 $\pi$), which
  correpond to the fastest possible migration towards pressure maxima
  (right) for runs IsoT100K2, FLD100K2, NIsoT100K2 and Adia100K2 after
  27.9 orbits. The legend for the different curves is in the right panel.}
\label{fig:sopt}
\end{figure}

The velocity, density and temperature of the gas flow in the circumplanetary disk can determine whether rocky satellites of giant
planets may form or not. Based on the strong sub-keplerianity of the
gas around their Jupiter-mass planet, \citet{KK06}
concluded that satellites could not form in their radiative transfer simulations. 

In Figure \ref{fig:sopt}, right panel, we show the particle size for which $T_s/T_{orb}$=1/(2 $\pi$) where $T_s$ is the stopping time in the Epstein regime and $T_{orb}$ is local the orbital time.
For any location in the circumplanetary disk grains whose size obeys the latter relation are the fastest to migrate towards pressure maxima 
(i.e. towards the planet in this case given that pressure 
decreases monotonically from the planet outwards). 
Much smaller particles will be entrained with the gas and migrate slowlier to the planet, while much larger boulders will follow 
perturbed keplerian orbits around the planet \citep{Weidenschilling77}. 
For isothermal and FLD runs the fastest migrating grains have sizes in the range 0.2 to 0.8~m, and the stopping time is
almost the same in both cases.
We note that our calculation is strictly valid only down to $0.15$ AU from the planet; at smaller distances
the assumption of the Epstein regime breaks down. 
For the adiabatic case, the Epstein regime is valid for bodies up  to 10~m in size because the density of the circumplanetary envelope is very low.
However, since the gas there is very hot, the fastest migrating grains are smaller than for the isothermal and FLD cases, 
having sizes in the range 0.06 to 0.1~m. 

Based on these results, we can expect that only very large boulders, i.e. those with sizes comfortably larger than indicated by
the curves in Figure~4, e.g. in the range $> 10-100$~m, may experience a slow enough radial 
decay to allow the assembly of satellites in the disk.
However, addressing whether or not satellites can really form will require to include grain growth and the Stokes 
drag regime, as well as self-consistent initial sizes and velocities for boulders entering the circumplanetary disk. 

\section{Conclusions}

We have shown the disk thermodynamics has an impact on all the most important aspects of disk-planet interaction for
Jupiter-mass planets. The way migration and accretion are affected is qualitatively consistent with the findings  
by \citet{Paardekooper06} for low-mass planets, although in that work the effect was quantitatively much stronger.

Migration is slowed down relative to isothermal runs because heating modifies the relative strength of the inner and outer spiral arm,
or stifles the spiral perturbations completely as in the case of adiabatic runs. This effect partially counterbalances the faster 
migration found in our isothermal disks relative to standard estimates that neglect viscosity, self-gravity and the motion of 
the planet. If the slower migration seen in radiative transfer runs persists over $> 100$ orbits one would expect the mean 
migration rate to differ from the estimates of standard type II migration by no more than a factor of 2, confirming the earlier 
conclusions reached by d'Angelo et al. (2003) using 2D simulations.
The mass transport towards the planet is hampered by the increasing pressure gradient in non-isothermal runs. This will have
to be investigated further by including a proper accretion model in the simulations \citep{d'Angelo03,KK06}. 


The different structure of the circumplanetary envelope depending
on disk thermodynamics that we find is qualitatively
in agreement with the results of \citet{KK06}, although our disks with flux-limited diffusion are less sub-keplerian than theirs.
Satellite formation is not easy in the thick circumplanetary disk, but it may still be possible provided that the solid component 
is dominated by large boulders.


\section{Acknowledgments}

The authors wish to thank H. Klahr, S.-J. Paardekooper and G. Mellema for fruitful discussions.
The simulations were run on the GONZALES cluster at ETH Zurich and on the ZBOX2 supercomputer 
at the University of Zurich.

\end{document}